\newcommand{\NPA}[3]{Nucl.\ Phys.\ A\ {\bf #1},\ #2 (#3)}

\newcommand{\PLB}[3]{Phys.\ Lett.\ B\ {\bf #1},\ #2 (#3)}

\newcommand{\PRL}[3]{Phys.\ Rev.\ Lett.\ {\bf #1},\ #2 (#3)}

\newcommand{\PRC}[3]{Phys.\ Rev.\ C\ {\bf #1},\ #2 (#3)}
\newcommand{\PRD}[3]{Phys.\ Rev.\ D\ {\bf #1},\ #2 (#3)}

\newcommand{\diracslash}[1]{#1\llap{/\kern2pt}}

\newcommand{\be}{\begin{equation}}
\newcommand{\ee}{\end{equation}}
\newcommand{\bea}{\begin{eqnarray}}
\newcommand{\eea}{\end{eqnarray}}
\newcommand{\ba}[1]{\begin{array}{#1}}
\newcommand{\ea}{\end{array}}

\documentclass[prd,aps,floats,nofootinbib,tightenlines,showpacs]{revtex4}
\usepackage{epsfig,graphicx,pstricks}
\usepackage{psfrag}
\usepackage{color}
\usepackage{amsmath}
\usepackage{amsfonts}
\usepackage{amssymb}
\usepackage{textcomp}
\usepackage{multirow}
\usepackage{subfigure}

\begin{document}

\title {Dissipative properties of hot and dense hadronic matter in excluded volume hadron resonance gas model}
\author{Guru Prakash Kadam }
\email{guruprasad@prl.res.in}
\affiliation{Theory Division, Physical Research Laboratory,
Navrangpura, Ahmedabad 380 009, India}
\author{Hiranmaya Mishra}
\email{hm@prl.res.in}
\affiliation{Theory Division, Physical Research Laboratory,
Navrangpura, Ahmedabad 380 009, India}

\date{\today} 

\def\be{\begin{equation}}
\def\ee{\end{equation}}
\def\bearr{\begin{eqnarray}}
\def\eearr{\end{eqnarray}}
\def\zbf#1{{\bf {#1}}}
\def\bfm#1{\mbox{\boldmath $#1$}}
\def\hf{\frac{1}{2}}
\def\sl{\hspace{-0.15cm}/}
\def\omit#1{_{\!\rlap{$\scriptscriptstyle \backslash$}
{\scriptscriptstyle #1}}}
\def\vec#1{\mathchoice
        {\mbox{\boldmath $#1$}}
        {\mbox{\boldmath $#1$}}
        {\mbox{\boldmath $\scriptstyle #1$}}
        {\mbox{\boldmath $\scriptscriptstyle #1$}}
}

\begin{abstract}
  We estimate dissipative properties viz: shear and bulk viscosities of hadronic matter using relativistic Boltzmann equation in relaxation time approximation within ambit of excluded volume hadron resonance gas (EHRG) model.  We find that at zero  baryon chemical potential the shear viscosity to entropy ratio ($\eta/s$) decreases with temperature while at finite baryon chemical potential this ratio shows same behavior as a function of temperature but reaches close to Kovtun-Son-Starinets (KSS) bound. Further along chemical freezout curve, ratio $\eta/s$ is almost constant apart from small initial monotonic rise. This observation may have some relevance to the experimental finding that the differential elliptic flow of charged hadrons does not change considerably at lower center of mass energy.  We further find that bulk viscosity to entropy density ($\zeta/s$) decreases with temperature while this ratio has higher value at finite baryon chemical potential at higher temperature. Along freezout curve $\zeta/s$ decreases monotonically at lower center of mass energy and then saturates. 
\end{abstract}


\maketitle

\section{Introduction}
In the context of heavy ion collision (HIC) experiments, the study of transport coefficients like shear and bulk viscosities is very important since these coefficients govern the evolution of the non-equilibrium system  towards equilibrium state. In the course of nuclear collision the spatial anisotropy gets converted into momentum anisotropy of produced particles. The equilibration of this momentum anisotropy is governed by shear viscosity coefficient. Bulk viscosity relates the momentum flux with velocity gradient. Based on AdS/CFT duality authors in Ref.\cite{kss} showed that there is a lower bound to the value of shear viscosity to entropy ratio ($\eta/s$) for any fluid found in nature. The lower bound (known as KSS bound) has been set to the value $1/4\pi$. Indeed, the elliptic flow measurement at RHIC led to $\eta/s$ very close to this lower bound\cite{hirano}. This discovery spurt out intense interest in transport properties of quantum chromodynamics (QCD). Unlike shear viscosity, bulk viscosity is very rarely studied transport coefficient. It was believed that the bulk viscosity does not play any significant role in the hydrodynamic evolution of the matter created in HIC experiments. At very high temperature QCD shows conformal symmetry whence the bulk viscosity to entropy ratio ($\zeta/s$) is very small. But near QCD transition temperature ($T_{c}$), conformal symmetry breaking may be significant because the  trace anomaly, $(\epsilon-3P)/T^{4}$ shows peak as observed in lattice simulations\cite{hotQCD,borsonyi}. So one would expect to observe similar peak in $\zeta/s$ at $T_{c}$ as well. Indeed rise of bulk viscosity near $T_{c}$ has been observed in various effective models\cite{dobado,sasakiqp,sasakinjl,ellislet,karschkharzeev,finazzo,WPblk,jeonyaffe}.

In order to distinguish signs of quark gluon plasma from that of hadronic matter, it is of fundamental importance to know the transport coefficients of hadronic matter. But computing these coefficients of hadronic matter is rather difficult task. There have been various efforts to compute them using some approximate schemes like Boltzmann transport equation in relaxation time approximation and Chapman-Enscog approximation\cite{prakash,prakashwiranata,purnendu,khvorostukhin,greco,greinerprc,gorenstein,cpsingh,sghoshnjl,weise,sarkarghosh,wiranatakoch,wiranataprc,wiraprapurn,tawfik}. Some authors have used Green-Kubo formalism involving correlation functions and various effective models for hadronic interaction  to compute transport properties for hadronic matter\cite{greinerprl,gurunpa,gurumpla,sghosh,sghoshnucl}. Apart from these analytic computations shear and bulk viscosity coefficients have also been estimated using transport codes like ultra-relativistic quantum molecular dynamics (UrQMD) transport codes\cite{demir} and parton hadron string dynamics (PHSD) codes\cite{phsd}. 

In present work, we estimate shear and bulk viscosities of hadronic matter using relativistic Boltzmann equation in relaxation time approximation within ambit of (thermodynamically consistent) excluded volume hadron resonance gas (EHRG) model\cite{ehrgrishke,ehrgclaymans,yen,gorengreinerjpg,gorengreinerprc,gorensteinplb,gorensteinjpg,yengorenprc}. We assume that all hadrons have uniform hard core radius. Thus our system under study is just a gas of hard sphere hadrons. As a good approximation we will use averaged partial relaxation time to compute viscosities. We work only in Boltzmann approximation i.e treat hadrons as classical Boltzmann particles which is rather good approximation at moderate temperature and take into account only elastic scattering processes.

We will organize the paper as follows. In section II we will briefly describe the thermodynamically consistent excluded volume hadron resonance gas model. In section III we will give brief derivation of shear and bulk viscosities for multicomponent hadronic matter in relaxation time approximation and then simplify them for the case of gas of hard spheres assuming averaged partial relaxation time for each hadron species. In section IV we will present the results  and discuss the implications of these results in the context of relativistic heavy ion collision. Finally we will summarize and conclude in section IV. 
 
\section{Excluded volume hadron resonance gas model}
The only mathematically consistent statistical ensemble to describe a physical system where the particle number is not conserved, which is the case for relativistic heavy ion collision, is grand canonical ensemble. The knowledge of partition function ($Z$) of an ensemble is sufficient to obtain all the thermodynamical quantities of such a system. The pressure ($P$) of an ideal gas can be obtained from $Z$ as

\be
P_{ideal}=T \lim_{V \rightarrow \infty}\frac{ln Z_{ideal}(T,\mu,V)}{V}
\ee
where $T$ is temperature, $\mu$ chemical potential and $V$ is volume of the system.
In thermodynamically consistent excluded volume formulation where the short range repulsive interaction has been taken into account via van der Waals (VDW) correction to the volume  one obtains the transcendental equation for the pressure as\cite{ehrgrishke}

\be
P(T,\mu)=P_{ideal}(T,\tilde{\mu})
\ee
where $\tilde{\mu}=\mu-vP(T,\mu)$ is an effective chemical potential with $v=\frac{4}{3}4\pi r_{h}^{3}$ as the parameter corresponding to proper volume of the particle. 
In classical Boltzmann approximation this prescription is equivalent to additional factor of $exp(-vP/T)$ to the pressure. Thus the pressure  in excluded volume hadron resonance gas in Boltzmann approximation is

\be
P(T,\mu)=exp(-vP(T,\mu)/T)P_{ideal}(T,\mu)
\ee
where $P_{ideal}$ in Boltzmann approximation can be written as

\be
P_{ideal}(T,\mu)=\sum_{a}\frac{g_{a}}{2\pi^{2}}m_{a}^{2}T^{2}K_{2}(\frac{m_{a}}{T})cosh(\frac{\mu}{T})
\ee
where $g_{a}$ is degeneracy of $a^{th}$ hadron species and $K_{2}$ is modified Bessel's function. We note that this expression for pressure is self consistent for given T and $\mu$ and the pressure of interacting gas is always smaller than that of ideal gas. Extending HRG model by VDW type correction leads to well known suppression of particle number densities consistent with heavy ion collision experiments\cite{yen}. 
Other thermodynamical quantities can be readily obtained from pressure, viz: Baryon number density $n_{B}=\partial P(T,\mu)/\partial \mu$, entropy density $s(T,\mu)=\partial P(T,\mu)/\partial T$, energy density $\varepsilon(T,\mu)=Ts(T,\mu)-P(T,\mu)+\mu n_{B}(T,\mu)$, sound speed $C_{s}^{2}(T,\mu)=dP(T,\mu)/d\varepsilon(T,\mu)$.

\section{Dissipative properties in relaxation time approximation}
In this section we will follow method given in Ref.\cite{gavin} to calculate transport properties in relaxation time approximation. Fundamental equation of the kinetic theory is Boltzmann transport equation given by
\be
\frac{\partial f_{p}}{\partial t}+ v_{p}^{i}\frac{\partial f_{p}}{\partial x^{i}}=I\{f_{p}\}
\label{boltzeqn}
\ee
$\vec v_{p}=\vec p/E_{p}$ is single particle velocity and $I\{f_{p\}}$ is collision integral and gives the rate of change of distribution  function $f_{p}$ due to collisions. $f_{p}$ is in general non-equilibrium distribution function.

Relaxation time approximation assumes that the collisions always bring the system to the local equilibrium exponentially with relaxation time which is of order of collision time. Thus in this approximation collision integral can be written as
\be
I\{f_{p}\}\backsimeq -\frac{(f_{p}-f_{p}^{0})}{\tau(E_{p})}
\label{rtapprx}
\ee
where $\tau(E_{p})$ is called relaxation time or collision time which in general depends on energy of the particle and $f_{p}^{0}$ is equilibrium distribution function  given by
\be
f_{p}^{0}=\frac{1}{exp\bigg(\frac{E_{p}-\vec p.\vec u-\mu}{T}\bigg)\pm1}
\label{distribution}
\ee
where $\vec u$ is fluid velocity and $\pm$ corresponds to fermions and bosons respectively. In hydrodynamical description of QCD matter shear and bulk viscosities enters in dissipative part ($T_{dissi}^{\mu\nu}$) of stress energy tensor.
\be
T^{\mu\nu}=T_{0}^{\mu\nu}+T_{dissi}^{\mu\nu}
\ee
where $T_{0}^{\mu\nu}$ is ideal part of stress tensor.
In the local lorentz frame dissipative part of stress energy tensor can be written as
\be
T_{dissi}^{ij}=-\eta\bigg(\frac{\partial u^{i}}{\partial x^{j}}+\frac{\partial u^{j}}{\partial x^{i}}\bigg)-(\zeta-\frac{2}{3}\eta)\frac{\partial u^{i}}{\partial x^{j}}\delta^{ij}
\label{dissi}
\ee

In terms of distribution function $T_{dissi}^{ij}$ can be written as
\be
T_{dissi}^{ij}=\int \frac{d^{3}p}{(2\pi)^{3}}p^{i}p^{j}\delta f_{p}
\label{dissi1}
\ee
where $\delta f_{p}$ is the deviation of the distribution function from equilibrium which governs the dissipative properties of the system. From Eq. (\ref{boltzeqn}) and Eq. (\ref{rtapprx}) we get
\be
\delta f_{p}=-\tau(E_{p})\bigg(\frac{\partial f_{p}^{0}}{\partial t}+ v_{p}^{i}\frac{\partial f_{p}^{0}}{\partial x^{i}} \bigg)
\label{deltadistr}
\ee
Assuming steady flow of the form $u^{i}=(u_{x}(y),0,0)$ and space-time independent temperature, Eq. (\ref{dissi}) simplifies to $T^{xy}=-\eta\partial u_{x}/\partial y$. From Eq. (\ref{dissi1}) and Eq. (\ref{deltadistr}) we get (using Eq.(\ref{distribution}) with $\mu=0$) 
\be
T^{xy}=\bigg\{-\frac{1}{T}\int\frac{d^{3}p}{(2\pi)^{3}}\tau(E_{p})\bigg(\frac{p_{x}p_{y}}{E_{p}}\bigg)^{2}f_{p}^{0}\bigg\}\frac{\partial u_{x}}{\partial y}
\ee

Thus coefficient of shear viscosity for a single component of hadronic matter is finally given by
\be
\eta=\frac{1}{15T}\int\frac{d^{3}p}{(2\pi)^{3}}\tau(E_{p})\frac{p^{4}}{E_{p}^{2}}f_{p}^{0}
\ee

Bulk viscosity is related to the dissipation in the system when it is uniformly compressed.  Taking trace of Eq. (\ref{dissi}) we get
\be
(T_{dissi})^{i}_{i}=-3\zeta \frac{\partial u^{i}}{\partial x^{i}}
\label{bulkdissi1}
\ee
Also from Eq. (\ref{dissi1}) and Eq. (\ref{deltadistr}) we get
\be
(T_{dissi})^{i}_{i}=-\int\frac{d^{3}p}{(2\pi)^{3}}\tau(E_{p})\frac{p^{2}}{E_{p}}\bigg(\frac{\partial f_{p}^{0}}{\partial t}+ v_{p}^{i}\frac{\partial f_{p}^{0}}{\partial x^{i}} \bigg)
\label{bulkdissi2}
\ee
Using energy momentum conservation law $\partial_{\mu}T^{\mu\nu}=0$, together with Eq. (\ref{bulkdissi1}) and Eq. (\ref{bulkdissi2})  one can arrive at\cite{gavin} 
\be
\zeta=\frac{1}{T}\int \frac{d^{3}p}{(2\pi)^{3}}\tau(E_{p}) f^{0}_{p}\bigg[E_{p}C_{n_{B}}^{2}-\frac{p^{2}}{3E_{p}}\bigg]^{2}
\ee
where $C_{n_{B}}^{2}=\frac{\partial P}{\partial\varepsilon}|_{n_{B}}$, is speed of sound at constant baryon density.

Thus for multicomponent hadron gas at finite chemical potential shear and bulk viscosities can be written as
\be
\eta=\frac{1}{15T}\sum_{a}\int \frac{d^{3}p}{(2\pi)^{3}}\frac{{p}^{4}}{E_{a}^{2}}
({\tau}_{a}f_{a}^{0}+{\bar\tau}_{a}\bar f_{a}^{0})
\label{shearmulti}
\ee

\bearr
\zeta&=&\frac{1}{T}\sum_{a}\int \frac{d^{3}p}{(2\pi)^{3}}\bigg\{\tau_{a} f^{0}_{a}\bigg[E_{a}C_{n_{B}}^{2}+\bigg(\frac{\partial P}{\partial n_{B}}\bigg)_{\varepsilon} -\frac{ p^{2}}{3E_{a}}\bigg]^{2}\nonumber\\&+&\bar\tau_{a} \bar f^{0}_{a}\bigg[E_{a}C_{n_{B}}^{2}-\bigg(\frac{\partial P}{\partial n_{B}}\bigg)_{\varepsilon}-\frac{ p^{2}}{3E_{a}}\bigg]^{2}\bigg\}
\label{blkmulti}
\eearr
where  $E_{a}^{2}=p^{2}+m_{a}^{2}$ and $(\partial P/\partial n_{B})_{\varepsilon}=n_{B}/(\partial n_{B}/\partial \mu)+C_{n_{B}}^{2}T^{2}\partial (\mu/T)/\partial T$. In above expressions bar stands for contribution of antiparticles.

  Energy dependent relaxation time is defined by expression
 \be
 \tau^{-1}(E_{a})=\sum_{bcd}\int\frac{d^{3}p_{b}}{(2\pi)^{3}}\frac{d^{3}p_{c}}{(2\pi)^{3}}\frac{d^{3}p_{d}}{(2\pi)^{3}}W(a,b\rightarrow c,d)f_{b}^{0}
 \label{tdrext}
 \ee
where the transition rate $W(a,b\rightarrow c,d)$ is defined by
\be
W(a,b\rightarrow c,d)=\frac{(2\pi)^{4}\delta(p_{a}+p_{b}-p_{c}-p_{d})}{2E_{a}2E_{b}2E_{c}2E_{d}}\mid \mathcal{M}\mid^{2}
\ee
with $\mid \mathcal{M}\mid$ being transition amplitude. In the center of mass frame  Eq. (\ref{tdrext}) can be simplified as
\be
\tau^{-1}(E_{a})=\sum_{b}\int\frac{d^{3}p_{b}}{(2\pi)^{3}}\sigma_{ab}\frac{\sqrt{S-4m^{2}}}{2E_{a}2E_{b}}f_{b}^{0}\equiv\sum_{b}\int\frac{d^{3}p_{b}}{(2\pi)^{3}}\sigma_{ab}v_{ab}f_{b}^{0}
\label{relx}
\ee
where $v_{ab}$ is relative velocity and $\sqrt{S}$ is center of mass energy. $\sigma_{ab}$ is the total scattering cross section for the process, $a(p_{a})+b(p_{b})\rightarrow a(p_{c})+b(p_{d})$.

For the simplicity we can use averaged relaxation time ($\tilde \tau$) which is rather a good approximation as energy dependent relaxation time\cite{moroz}. One can obtain $\tilde \tau$ as follows.  Averaging over $f_{a}^{0}$ Eq. (\ref{relx}) becomes
\be
\frac{\int\frac{d^{3}p_{a}}{(2\pi)^{3}}\tau^{-1}(E_{a})f_{a}^{0}}{\int\frac{d^{3}p_{a}}{(2\pi)^{3}}f_{a}^{0}}=\sum_{b}\frac{\int\frac{d^{3}p_{a}}{(2\pi)^{3}}\frac{d^{3}p_{b}}{(2\pi)^{3}}\sigma_{ab}v_{ab}f_{a}^{0}f_{b}^{0}}{\int\frac{d^{3}p_{a}}{(2\pi)^{3}}f_{a}^{0}}
\label{relx1}
\ee
Thus averaged partial relaxation time is given by
\be
{\tilde\tau}_{a}^{-1}=\sum_{b}n_{b}\langle\sigma_{ab}v_{ab}\rangle
\label{relxaverage}
\ee
where $n_{b}=\int\frac{d^{3}p_{b}}{(2\pi)^{3}}f_{b}^{0}$ is the number density of $b^{th}$ hadronic species.

In this work we will use equilibrium Maxwell-Boltzmann distribution (in the local rest frame) given by
\be
f_{a}^{0}=exp\bigg(-\frac{E_{a}-\mu}{T}\bigg)
\label{MB}
\ee

The thermal average of total cross section times relative velocity i.e  $\langle\sigma v\rangle$ for the scattering of hard sphere particles (having constant cross section, $\sigma$) of the same species at zero baryon density can be calculated as follows\cite{cannoni,gelmini}. With Maxwell-Boltzmann distribution $f^{0}(E)=exp(-E/T)$, the thermal average $\langle\sigma v\rangle$ for the process $a(p_{a})+a(p_{b})\rightarrow a(p_{c})+a(p_{d})$ can be written as
 \be
 \langle\sigma_{ab} v_{ab}\rangle=\frac{\sigma \int d^{3}p_{a}d^{3}p_{b} v_{ab}e^{-E_{a}/T}e^{-E_{b}/T}}{\int d^{3}p_{a}d^{3}p_{b}e^{-E_{a}/T}e^{-E_{b}/T}}
 \label{thermalave}
 \ee
 Momentum space volume elements $d^{3}p_{a}d^{3}p_{b}$ can be written in terms of scattering angle $\theta$ as
 \be
 d^{3}p_{a}d^{3}p_{b}=(4\pi)^{2}p_{a}p_{b}dE_{a}dE_{b}\frac{1}{2}cos\theta
 \ee
 Changing integration variables from $E_{a},E_{b}, \theta$ to $E_{-},E_{+},S$ we gets
  \be
 d^{3}p_{a}d^{3}p_{b}=2\pi^{2}E_{a}E_{b}dE_{-}dE_{+}dS
 \ee
 where $S=(p_{a}+p_{b})^{2}$ is usual Mandelstam variable and $E_{\pm}=E_{a}\pm E_{b}$.
 With this change in variables the integration region transform as
 
 \be
  E_{-}\leq\sqrt{1-\frac{4m^{2}}{S}} \sqrt{E_{+}^{2}-S}
 \ee
 with $E_{+}\geqslant \sqrt{S}$ and $S\geqslant 4m^{2}$.
 Thus numerator in Eq. (\ref{thermalave}) becomes
 \be
 \int d^{3}p_{a}d^{3}p_{b} v_{ab}e^{-E_{a}/T}e^{-E_{b}/T}=2\pi^{2}T\int dS \sqrt{S} (S-4m^{2})K_{1}(\sqrt{S}/T)
 \ee
 Similarly denominator of Eq. (\ref{thermalave}) can be evaluated as
 \be
 \int d^{3}p_{a}d^{3}p_{b}e^{-E_{a}/T}e^{-E_{b}/T}=[4\pi m^{2}T K_{2}(m/T)]^{2}
 \ee
 where $K_{n}$ is modified Bessel function of order n. Thus the thermal average becomes
 \be
 \langle\sigma_{ab} v_{ab}\rangle=\frac{\sigma}{8m^{4}TK_{2}^{2}(m/T)}\int_{4m^{2}}^{\infty}dS\sqrt{S}(S-4m^{2})K_{1}(\sqrt{S}/T)
 \ee
 For scattering between different species of the particles ($a(p_{a})+b(p_{b})\rightarrow a(p_{c})+b(p_{d})$) one can generalize above equation to get
 
\be
\langle\sigma_{ab}v_{ab}\rangle=\frac{\sigma}{8Tm_{a}^{2}m_{b}^{2}K_{2}(\frac{m_{a}}{T})K_{2}(\frac{m_{b}}{T})}\int_{m_{a}+m_{b}}^{\infty}dS\frac{[S-(m_{a}-m_{b})^{2}]}{\surd S}[S-(m_{a}+m_{b})^{2}]K_{1}(\surd S/T)
\label{thermalave1}
\ee
Computing the thermal averaged cross section as above, one can relate it to the relaxation time in Eq. (\ref{relxaverage}). The viscosities can then be calculated using  Eq. (\ref{shearmulti}) and Eq. (\ref{blkmulti}) once the thermodynamic quantities are estimated using EHRG model.

\section{Results and discussion}
To estimate the thermodynamical quantities using EHRG model we take all hadrons and their resonances with mass cut-off 2.25 GeV\cite{amseler} in the partition function. The only unknown parameter needed to compute viscosity coefficients is hadronic hard core radius $r_{h}$. On one hand in Ref.\cite{heppe} it has been argued that one can take uniform hard core radius equal to 0.3fm for all hadrons. This argument is based on the hard core repulsive interaction between nucleons. For mesons there is no detailed results for short range repulsive interaction. But one can set same hard core radius to all mesons based on similarity of charge radii compared to baryons and pion-nucleon phase shift energy dependence. On the other hand in Ref.\cite{bugaev} authors demonstrated that strangeness Horn behavior can be described in hadron resonance gas model without spoiling hadron yield fit if one set vanishing  hard core radius ($r_{h}$) for pions, $r_{h}=0.35$fm for kaons, 0.3fm for all other mesons and 0.5fm for all baryons. Based on all these arguments we take two values of $r_{h}$, 0.3fm and 0.5fm for all hadrons to compute viscosity coefficients. The cross section of the hadrons in terms of hadron hard core radius is given by $\sigma=4\pi r_{h}^{2}$.

\begin{figure}[h]
\vspace{-2.5cm}
\centering
 \includegraphics[width=8.5cm,height=8.5cm]{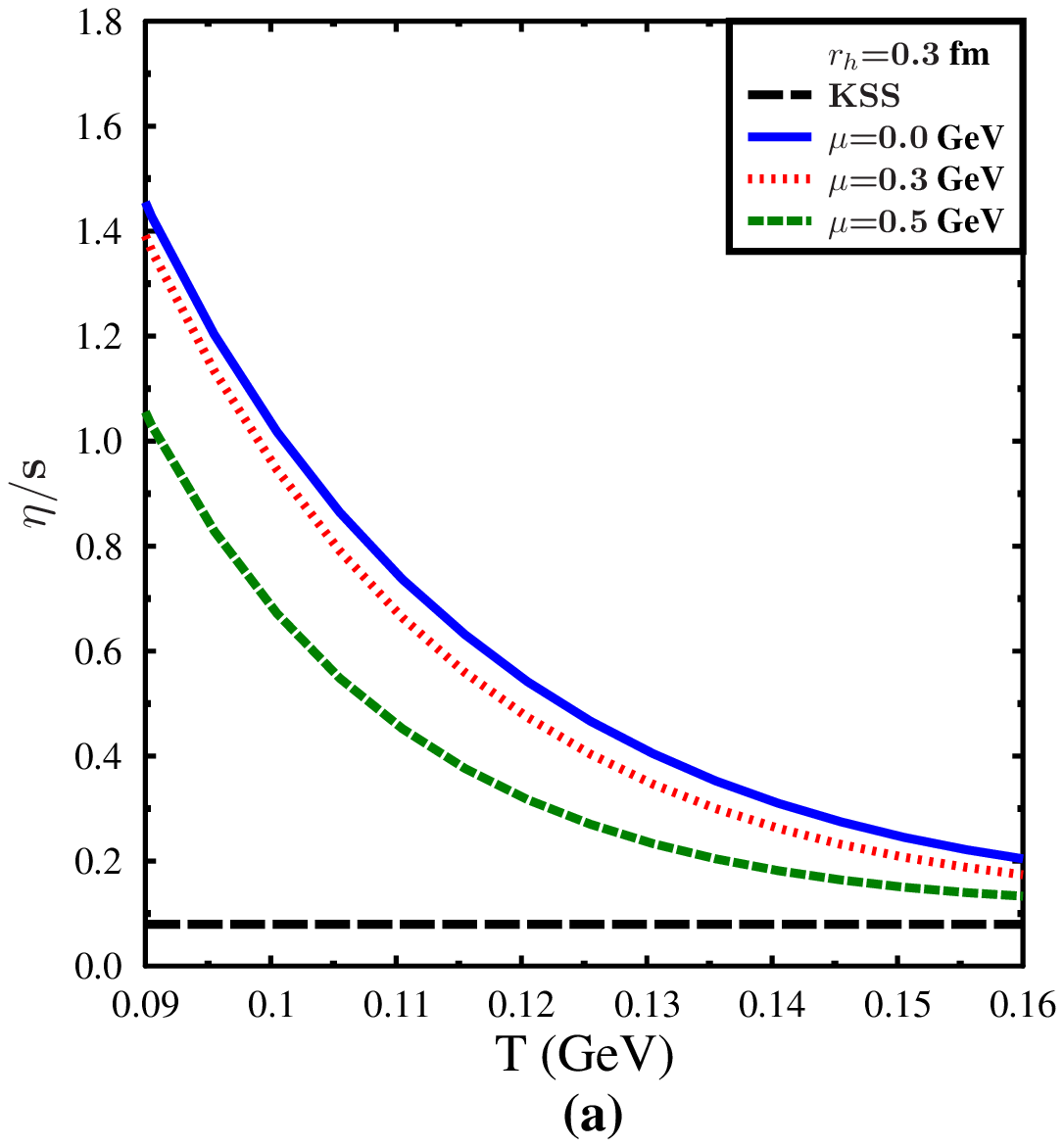}\\
  \includegraphics[width=8.5cm,height=8.5cm]{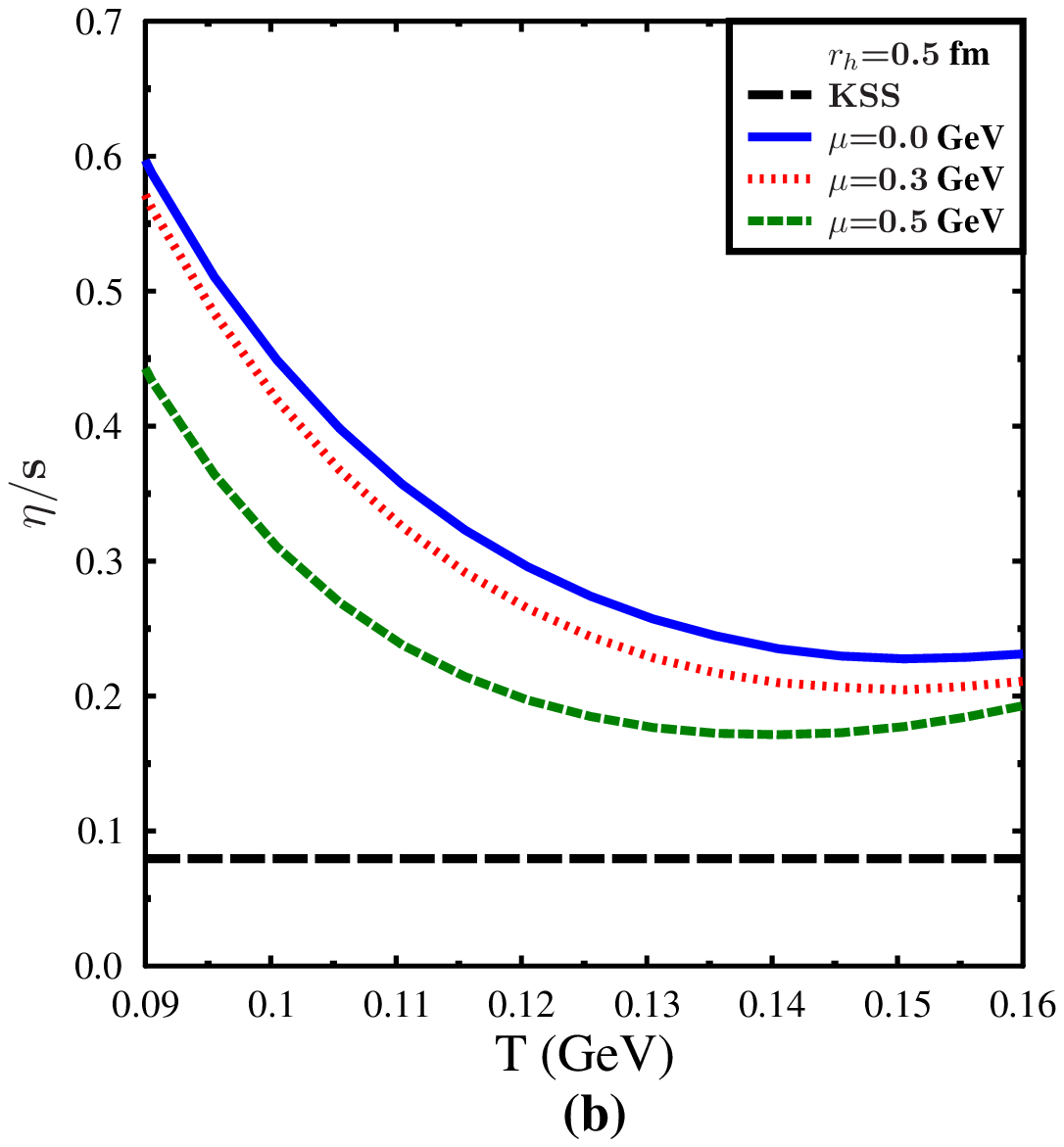}\\
   \includegraphics[width=8.5cm,height=8.5cm]{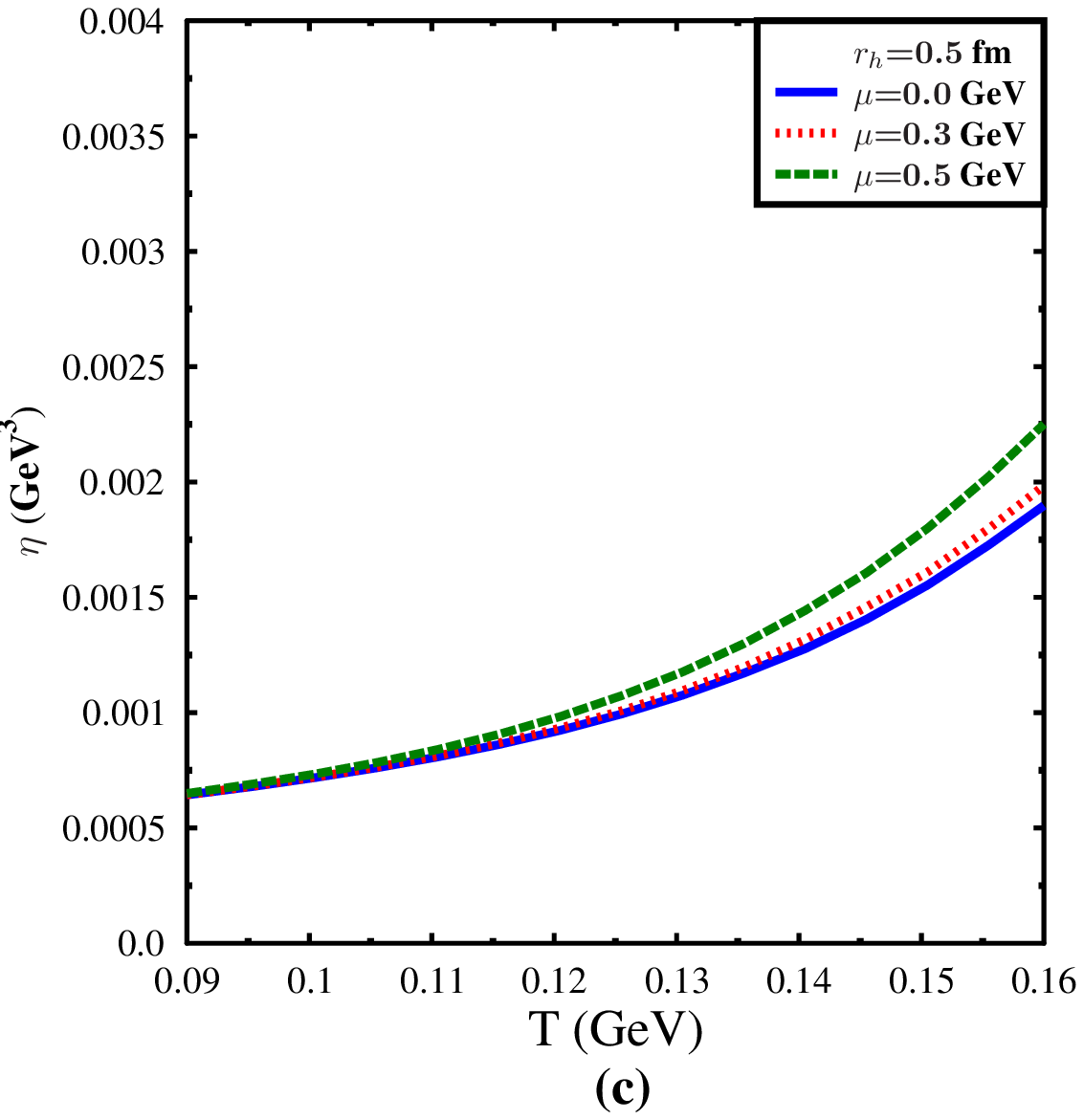}
  \caption{(Color online) Top  panel shows shear viscosity to entropy density ratio $\eta/s$ for $r_{h}=0.3$ fm as a function of temperature for different chemical potentials. Middle panel shows $\eta/s$ for $r_{h}=0.5$ fm. Bottom panel shows the shear viscosity  as a function of temperature at different chemical potentials for $r_{h}=0.5$ fm.} 
\label{etabis}
 \end{figure}

\begin{figure}[h]
\vspace{-0.4cm}
\begin{center}
 \includegraphics[width=8.5cm,height=8.5cm]{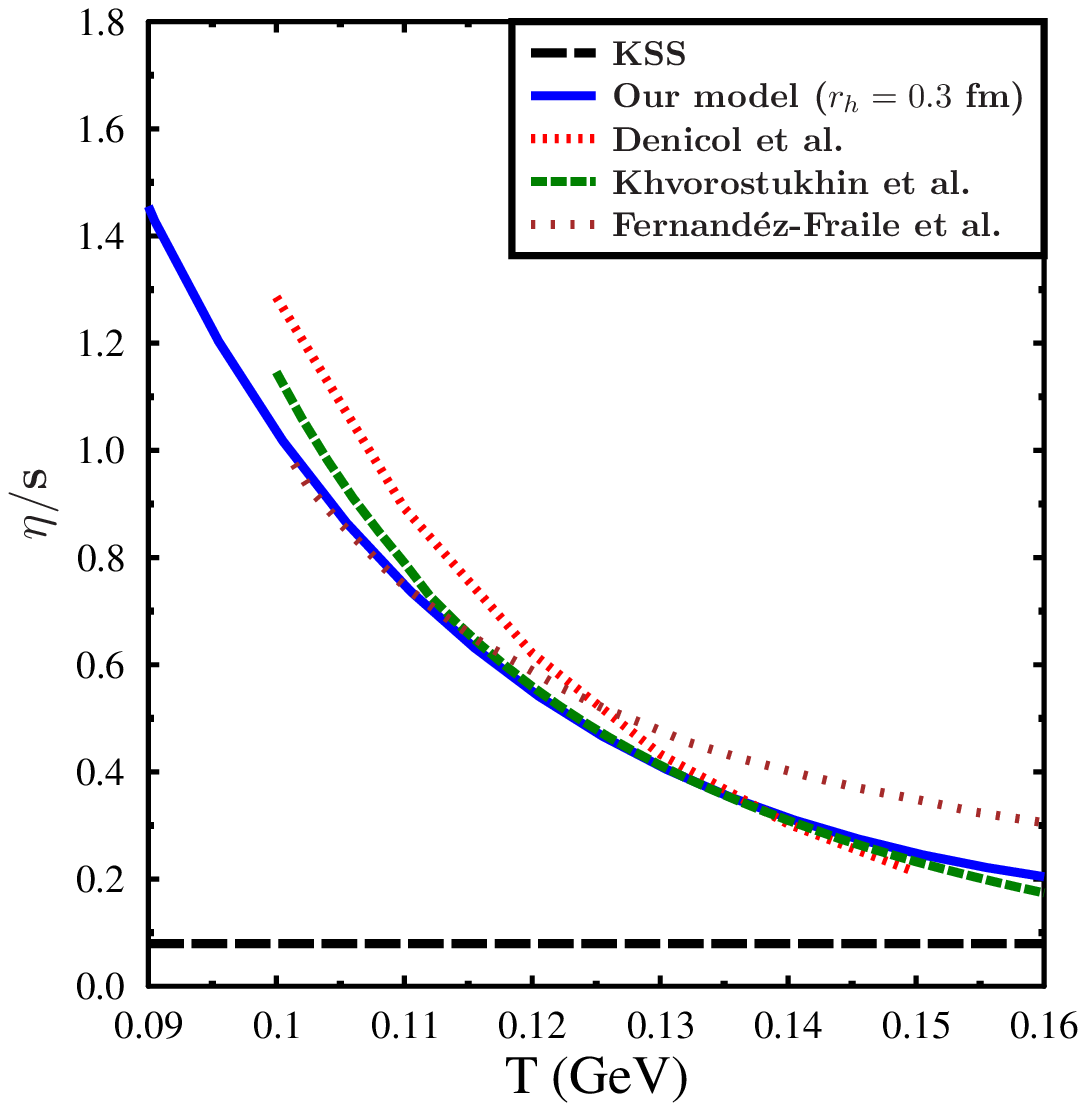} 
\caption{(Color online) Comparison of $\eta/s$  estimated in our model with the estimations of other models\cite{jeon,Khov,nicola} at zero chemical potential.
}
\label{etabis_compare}
\end{center}
 \end{figure}

\begin{figure}[h]
\vspace{-0.4cm}
\begin{center}
\begin{tabular}{c c}
 \includegraphics[width=8.5cm,height=8.5cm]{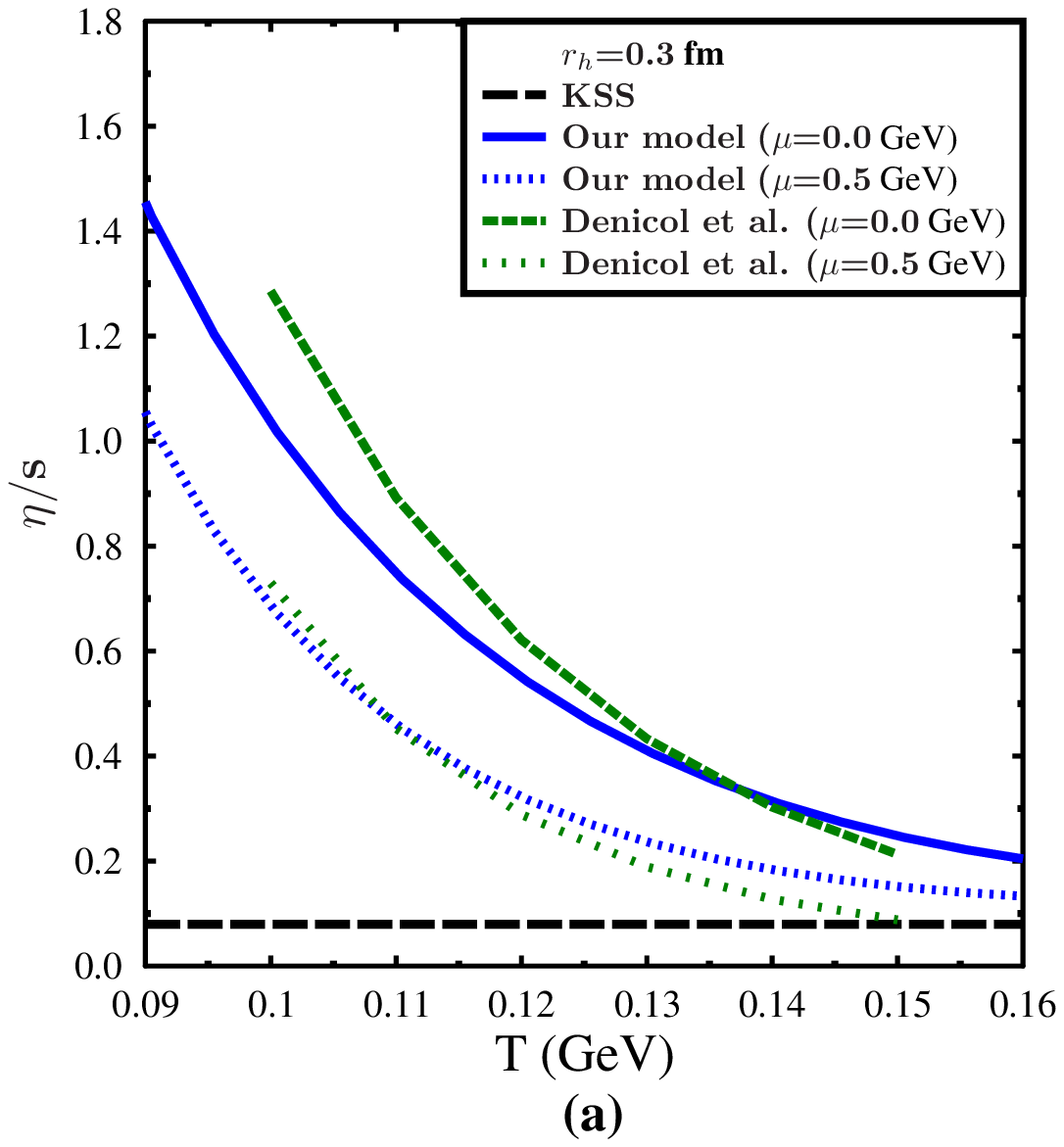}&
  \includegraphics[width=8.5cm,height=8.5cm]{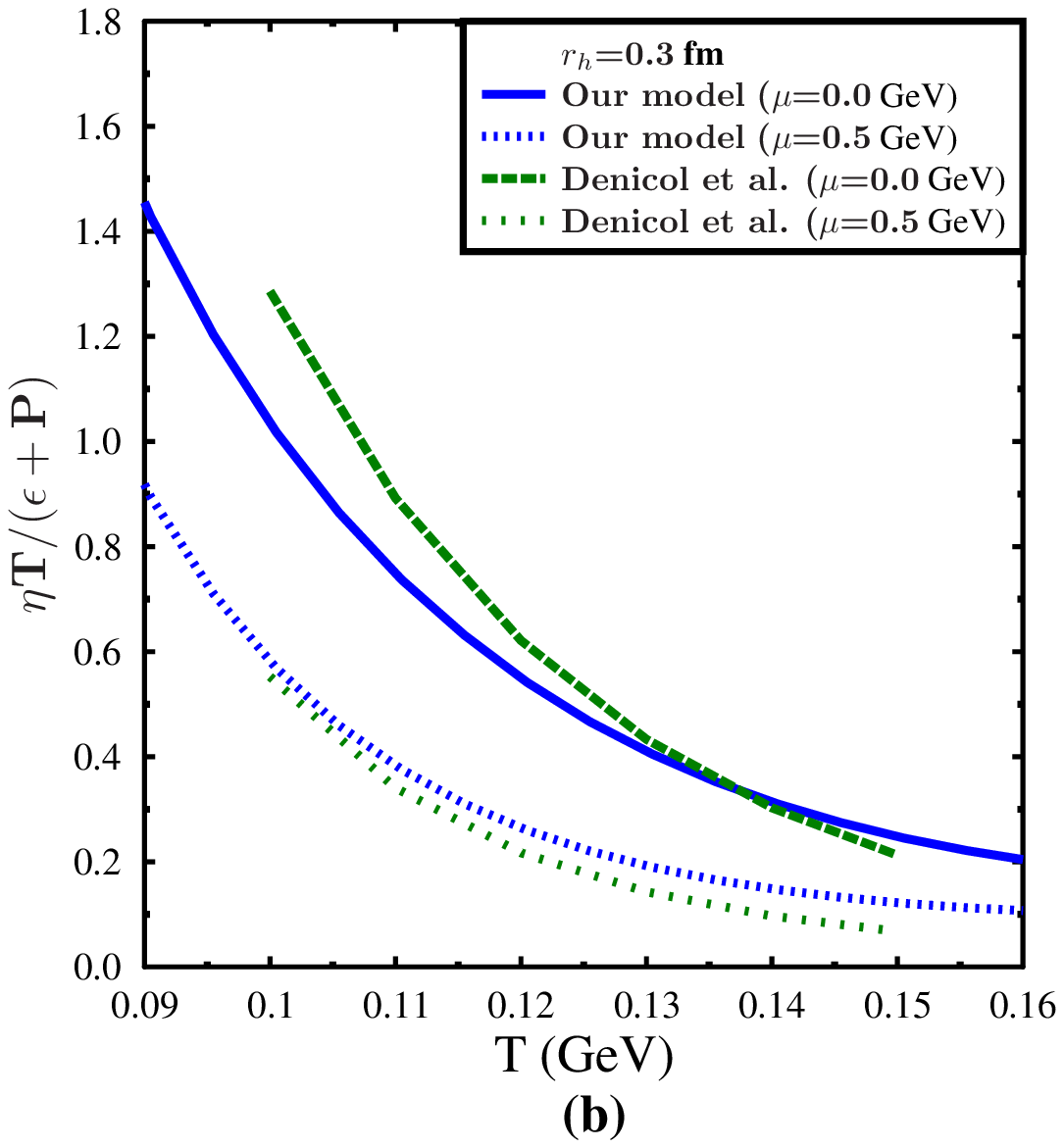}\\
  \end{tabular}
  \caption{(Color online) Left panel shows comparison of $\eta/s$ ($r_{h}=0.3$fm) estimated in our model with the estimations of Ref.\cite{jeon}. Right panel shows comparison of $\eta T/(\epsilon+P)$ with Ref\cite{jeon}. } 
\label{etabismu_compare}
  \end{center}
 \end{figure}

Results for shear viscosity coefficient is shown in Fig.(\ref{etabis}). The general behavior of $\eta/s$ is similar to that observed in Ref.\cite{gorenstein} where the authors considered EHRG within relativistic molecular kinetic theory unlike relaxation time approximation scheme used in this work. We also compare our results of $\eta/s$ with other model calculations like Chapman-Enscog theory\cite{jeon}, scaling
hadron masses and couplings (SHMC) model\cite{Khov} and chiral perturbation theory\cite{nicola} at zero baryon chemical potential as shown in Fig.(\ref{etabis_compare}). We note that the general behavior of $\eta/s$ is in conformity with these models. We also note that at low temperature ($\sim0.120$GeV) where the pions are dominating degrees of freedom, our results matches with Ref.\cite{nicola} where the authors estimates $\eta/s$ for the gas of pions using chiral perturbation theory while at high temperature (above 0.120 GeV) our results matches with Ref.\cite{Khov} where the authors estimated this ratio in SHMC model for hadronic matter.    Further we observe that at finite chemical potential, although the general behavior of the  ratio is similar as a function of temperature, ratio is smaller than that at $\mu=0$GeV and approaches closer to KSS bound.  Thus finite baryon chemical potential significantly affect $\eta/s$. Although the shear viscosity itself increases with $\mu$ as shown in Fig.(\ref{etabis}c), decrease in ratio $\eta/s$ at finite $\mu$ is solely due to rapid increase in entropy density. This behavior of $\eta/s$ at finite baryon density is consistent with Ref.\cite{jeon} where the authors have estimated $\eta/s$ using Chapman-Enskog theory within hadron resonance gas model. We compare our results with the results of Ref.\cite{jeon} at zero and finite $\mu$ as shown in Fig.(\ref{etabismu_compare}a). We note that the general behavior of $\eta/s$ is similar except the fact that the value approaches closer to KSS bound in Chapman-Enskog theory.

 It is important to note that at finite chemical potential, $\eta/s$ cannot be inferred as a measure of fluidity\cite{liao}. Also this ratio can be shown to violate KSS bound in kinetic theory.  Based on crude kinetic theory argument one can show that $\eta\approx\frac{1}{3}\sum_{a}(n\langle p\rangle \lambda)_{a}$, where n is number density, $\langle p\rangle$ is thermal momentum and $\lambda$ is mean free path. Kinetic theory is valid only for those gases for which mean free path is much smaller than the typical size of the system (L) i.e, $\lambda\ll L$ and for those gases for which $\lambda$ must be larger than inter particle spacing. Then uncertainty relation $\lambda \langle p\rangle\geq 1$ implies that there is a lower bound to shear viscosity, $\eta\gtrsim 2T^{3}$\cite{gyulassy}. Also in the non-relativistic limit one can show that the shear viscosity of the gas of hard spheres is independent of number of particle species\cite{prakashwiranata}.  On the other hand, entropy density  of the gas consisting multiple hadronic species (which also goes as $T^{3}$)  can be made very large so that ratio $\eta/s$ can be made arbitrarily small. In fact at sufficiently high chemical potential mixing entropy of multicomponent hadron gas overwhelms and hence ratio $\eta/s$ can go below KSS bound. This fact has been used in Ref.\cite{cohen} to give counterexample to KSS bound.  We might also like to mention here that KSS bound can be violated in certain field theories\cite{sinha,rebhan,mamo}.

\begin{figure}[h]
\vspace{-0.4cm}
\begin{center}
\begin{tabular}{c c}
 \includegraphics[width=8.5cm,height=8.5cm]{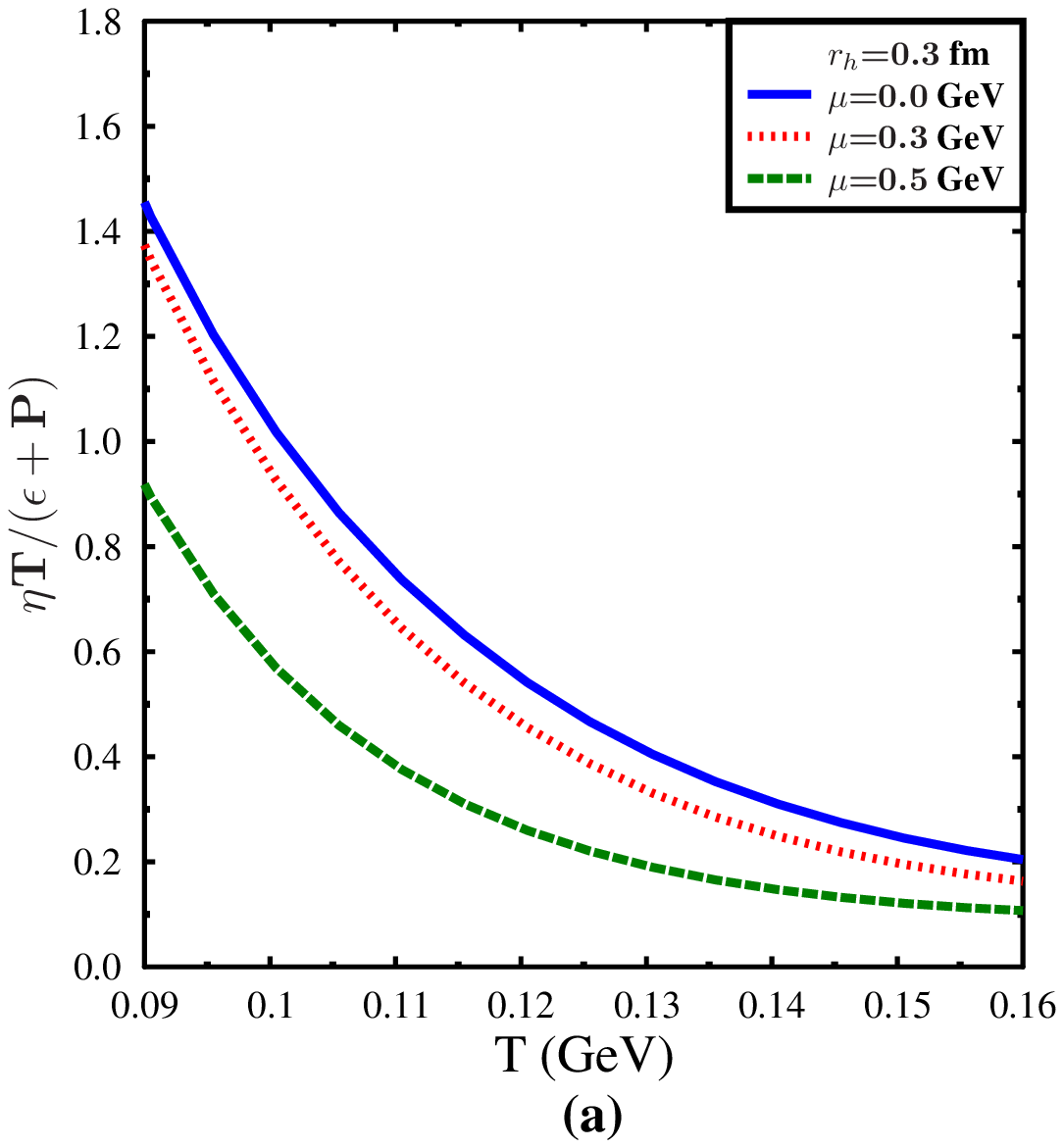}&
  \includegraphics[width=8.5cm,height=8.5cm]{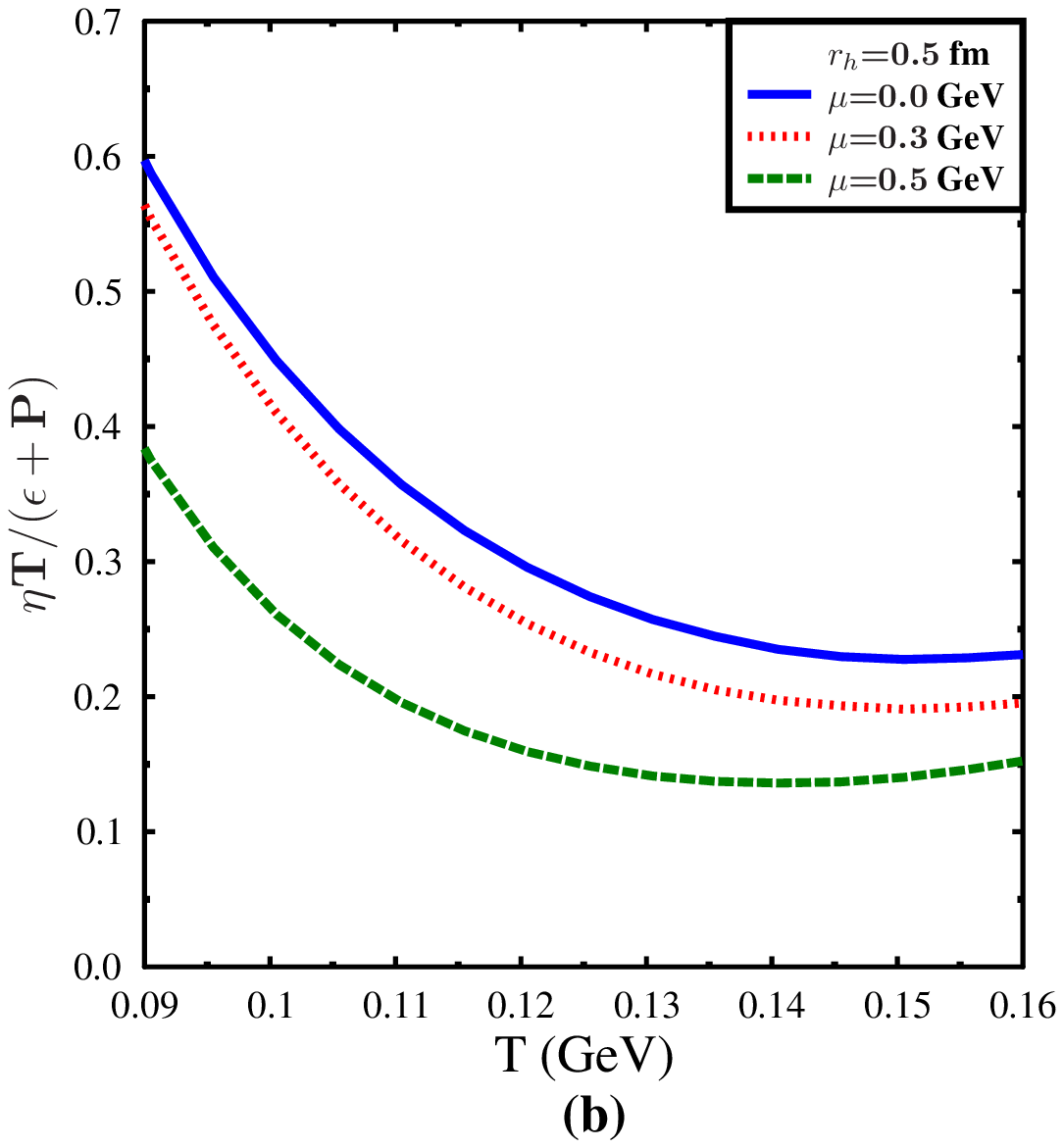}\\
  \end{tabular}
  \caption{(Color online) Left panel shows ratio $\eta T/(\epsilon+P)$ for $r_{h}=0.3$ fm as a function of temperature for different chemical potentials. Right panel shows $\eta T/(\epsilon+P)$ for $r_{h}=0.5$ fm.} 
\label{etabenth}
  \end{center}
 \end{figure}

At finite $\mu$, it is the ratio $\eta T/(\epsilon+P)$ which is correct measure of fluidity\cite{liao}. Quantity $\epsilon+P$ is called enthalpy and as per thermodynamical relation, $\epsilon+P=Ts+\mu n_{B}$, we note that at $\mu=0$ we get back $\eta/s$ as a fluidity measure. From Fig.(\ref{etabenth}) we note that effect of finite chemical potential is more pronounced in ratio  $\eta T/(\epsilon+P)$. This can again be attributed to rapid rise in enthalpy at finite $\mu$. The general behavior of the ratio $\eta T/(\epsilon+P)$ is again in conformity with Ref.\cite{jeon} as shown if Fig.(\ref{etabismu_compare}b) except the fact that for given chemical potential the ratio is smaller in Chapman-Enskog theory.
 
\begin{figure}[h]
\vspace{-2.5cm}
\centering
 \includegraphics[width=8.5cm,height=8.5cm]{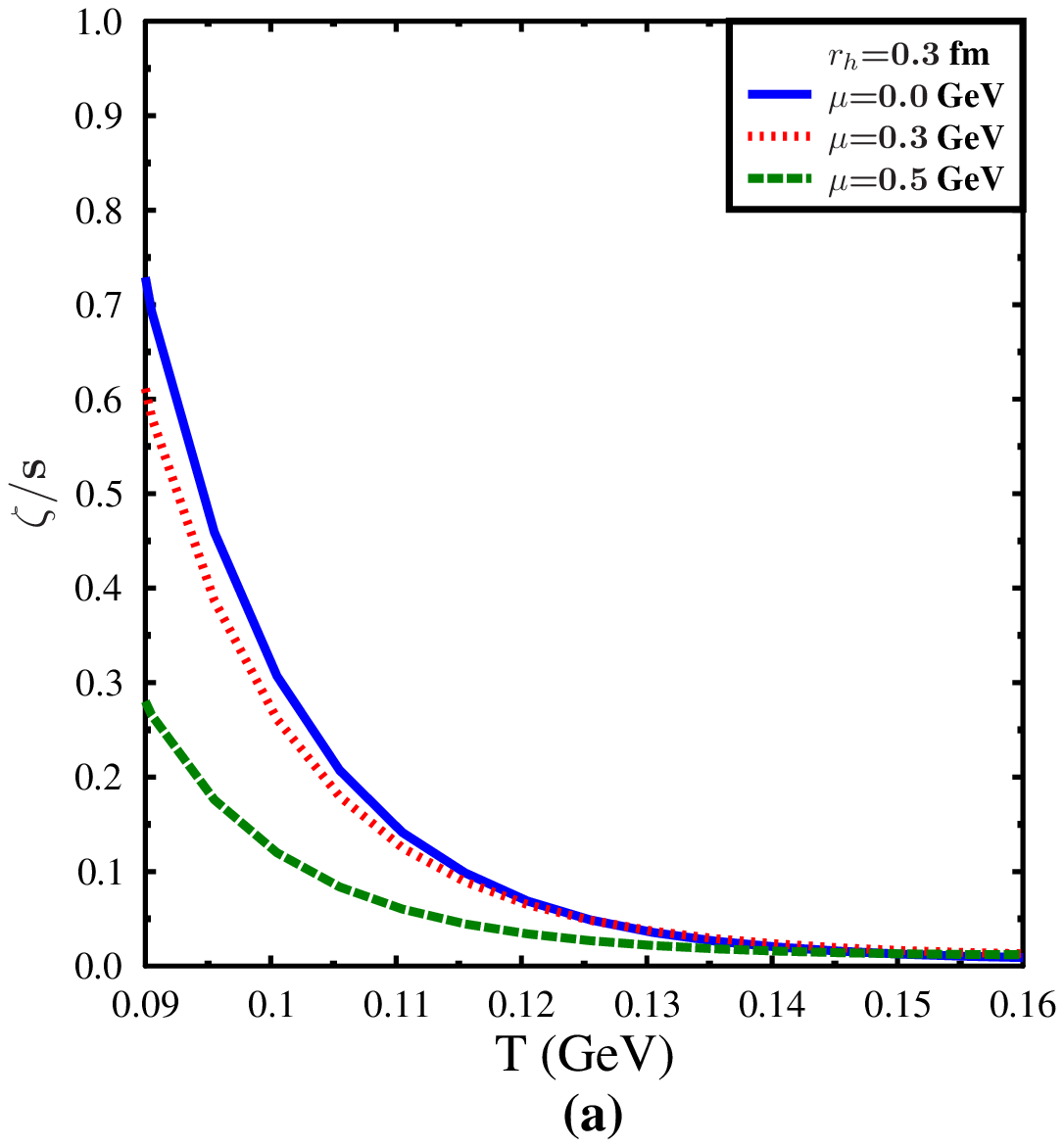}\\
  \includegraphics[width=8.5cm,height=8.5cm]{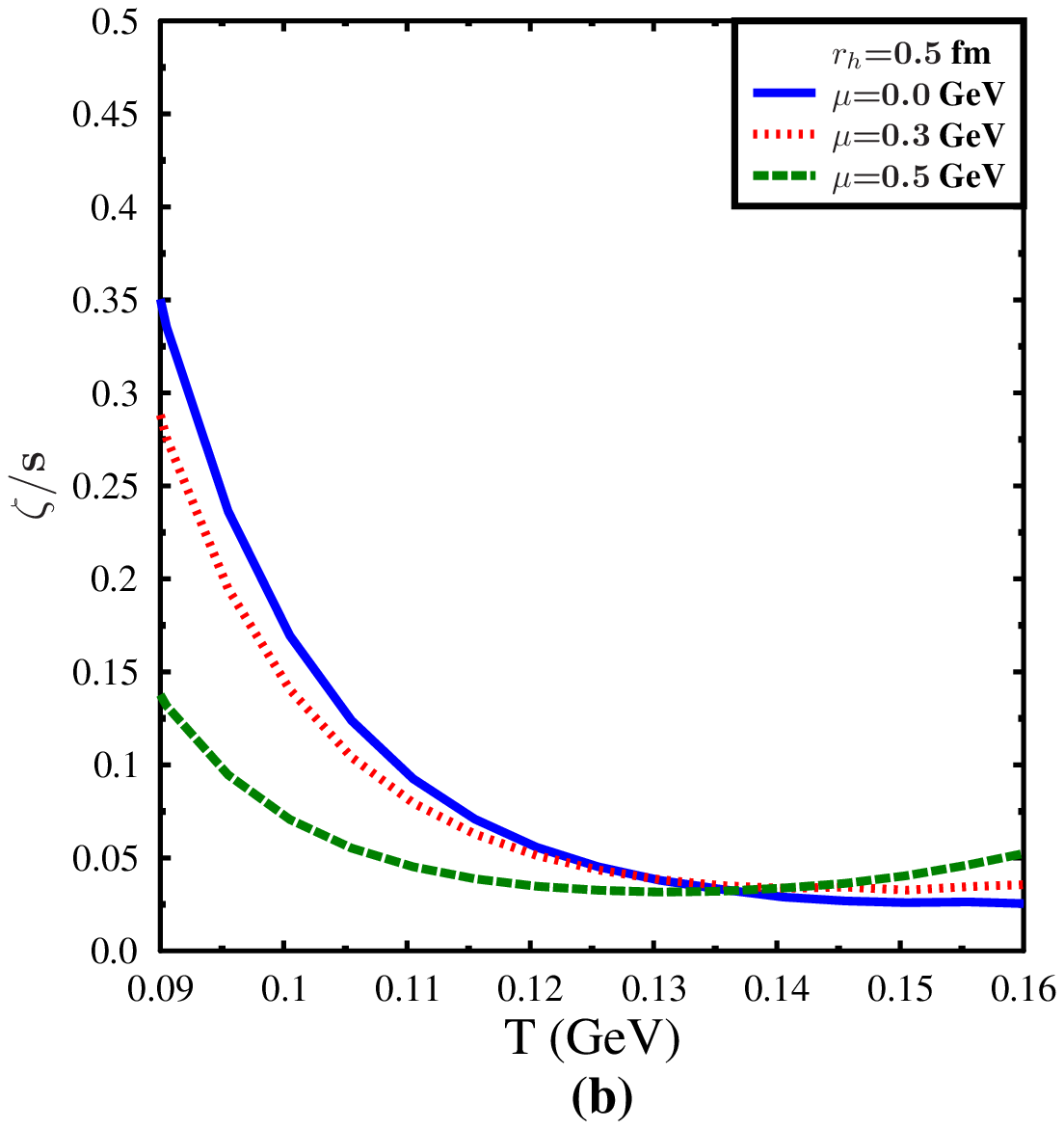}\\
   \includegraphics[width=8.5cm,height=8.5cm]{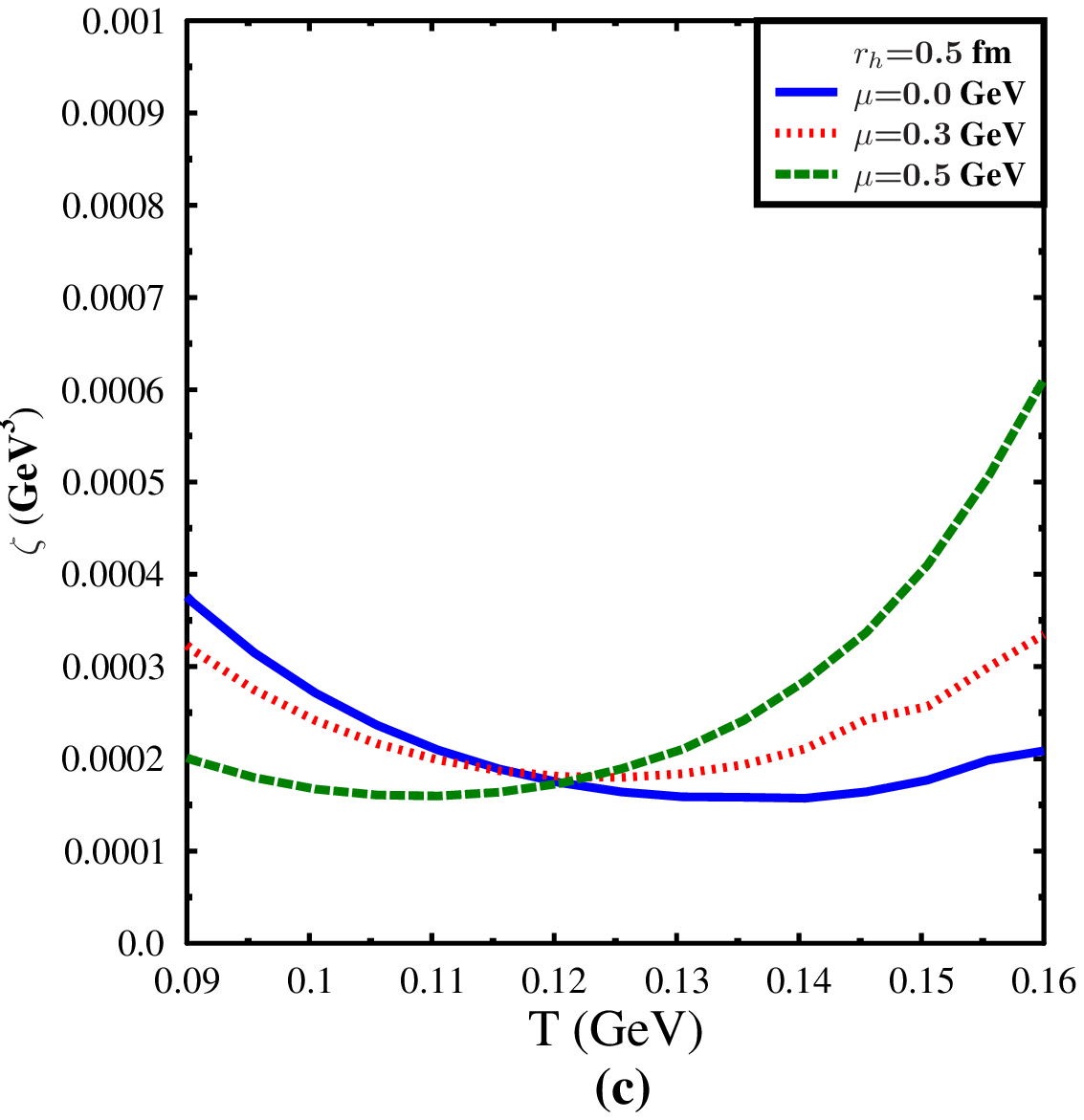}
  \caption{(Color online) Top panel shows bulk viscosity to entropy ratio $\zeta/s$ for $r_{h}=0.3$ fm as a function of temperature for different chemical potentials. Middle panel shows $\zeta/s$ for $r_{h}=0.5$ fm. Bottom panel shows  the bulk viscosity  as a function of temperature at different chemical potentials for $r_{h}=0.5$fm. } 
\label{zetabis}
 \end{figure}

\begin{figure}[h]
\vspace{-0.4cm}
\begin{center}
 \includegraphics[width=8.5cm,height=8.5cm]{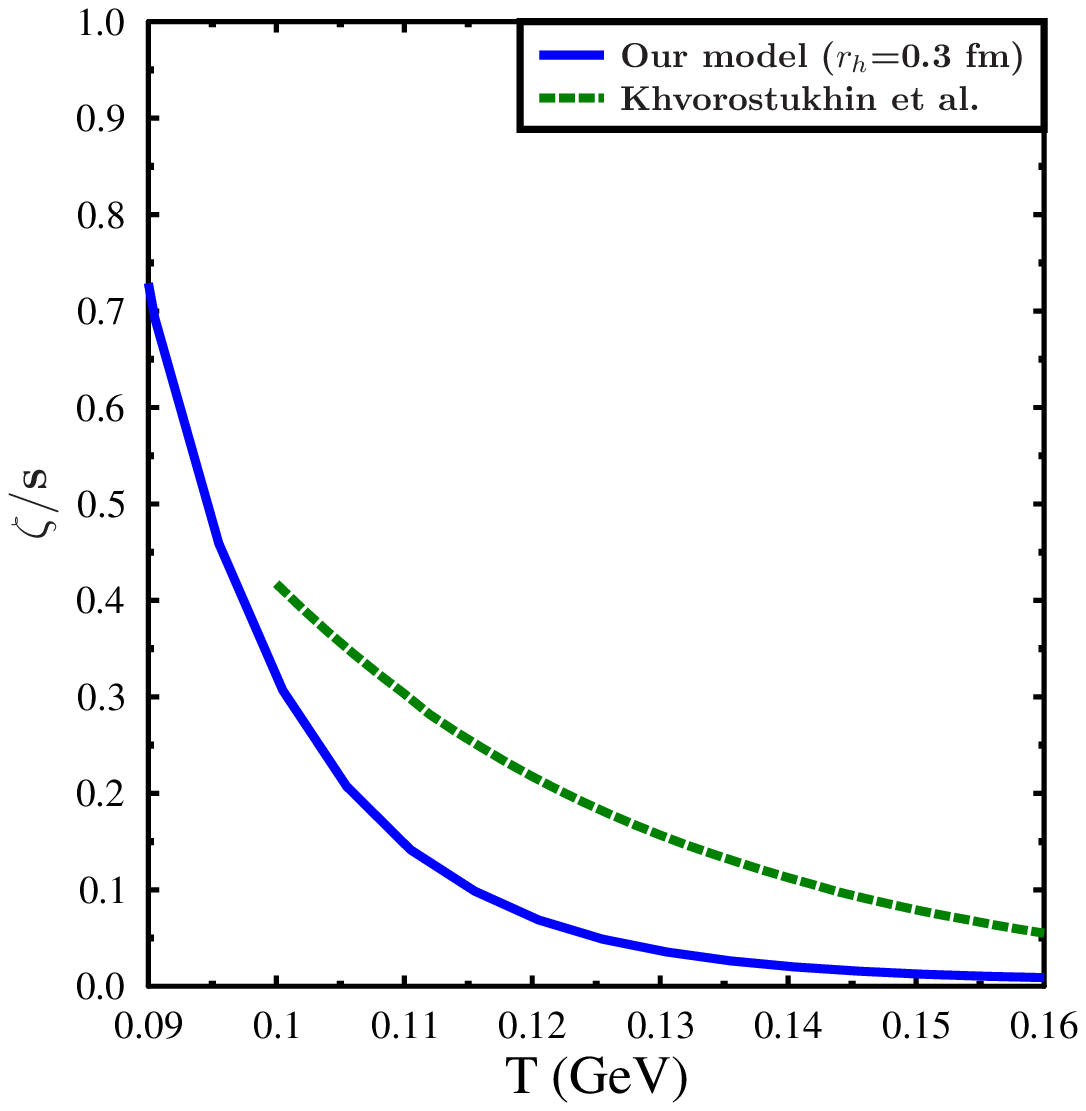} 
\caption{(Color online) Comparison of $\zeta/s$  estimated in our model with estimations using SHMC model\cite{Khov} at zero chemical potential.
}
\label{zetabis_compare}
\end{center}
 \end{figure}

 Fig.(\ref{zetabis}) shows results for the bulk viscosity. We note that ratio $\zeta/s$ decreases with temperature at zero chemical potential. As shown in Fig.(\ref{zetabis_compare}), the general behavior of $\zeta/s$ is  similar to that observed in Ref.\cite{Khov} where the authors estimated the bulk viscosity using SHMC model. At finite chemical potential although the ratio $\zeta/s$ decreases at low temperature, it increases in the window $T=0.120-0.160$ GeV. This is because bulk viscosity itself increases very rapidly in this window  as shown in Fig.(\ref{zetabis}c). This rise may be attributed to the explicit scale symmetry violation by finite chemical potential and hence the massive nucleon excitations which contribute more at finite baryon chemical potential\cite{gurunpa}. We might mention here that although the inelastic scattering processes needs to be taken into account for the precise estimation of the bulk viscosity\cite{moore}, authors in Ref.\cite{dobadojuan} showed that inelastic processes are irrelevant in the bulk viscosity computation at low and moderate temperatures. In Fig.(\ref{zetabis_compare}) we compare $\zeta/s$ estimated in our model with SHMC model\cite{Khov}. We note that our $\zeta/s$ curve vanishes faster at high temperature
 
\begin{figure}[h]
\vspace{-2.5cm}
\centering
 \includegraphics[width=8.5cm,height=8.5cm]{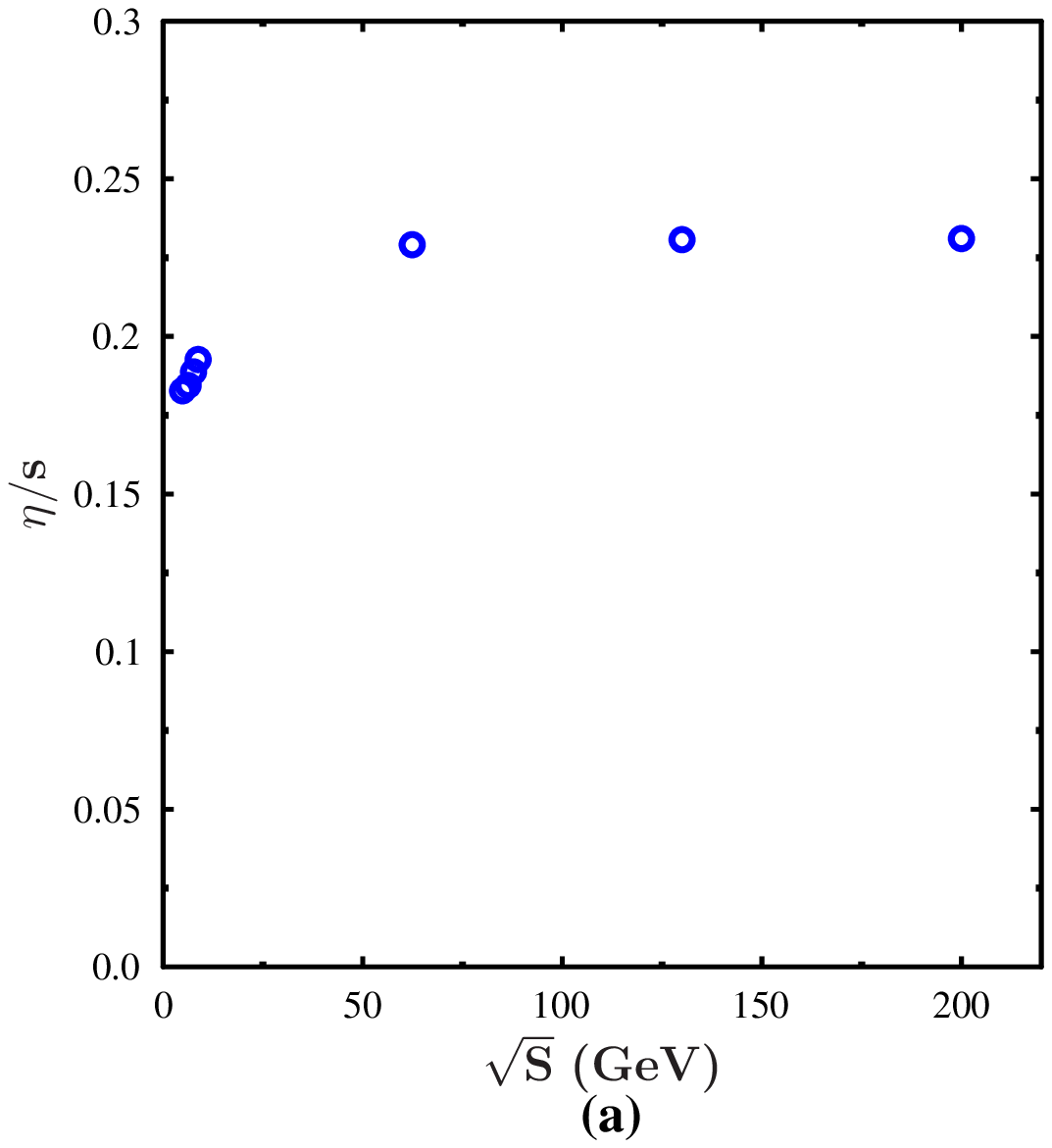}\\
  \includegraphics[width=8.5cm,height=8.5cm]{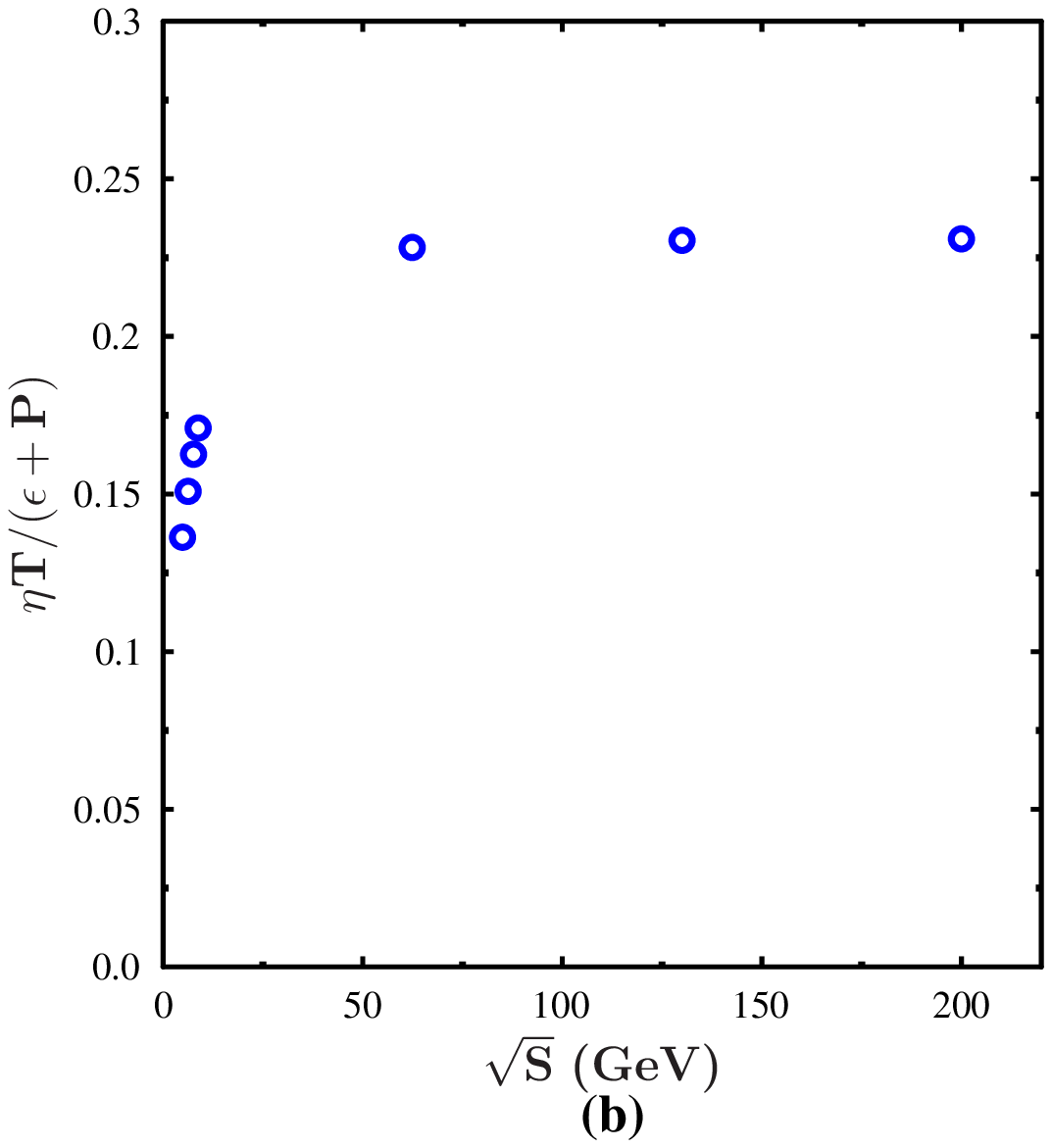}\\
  \includegraphics[width=8.5cm,height=8.5cm]{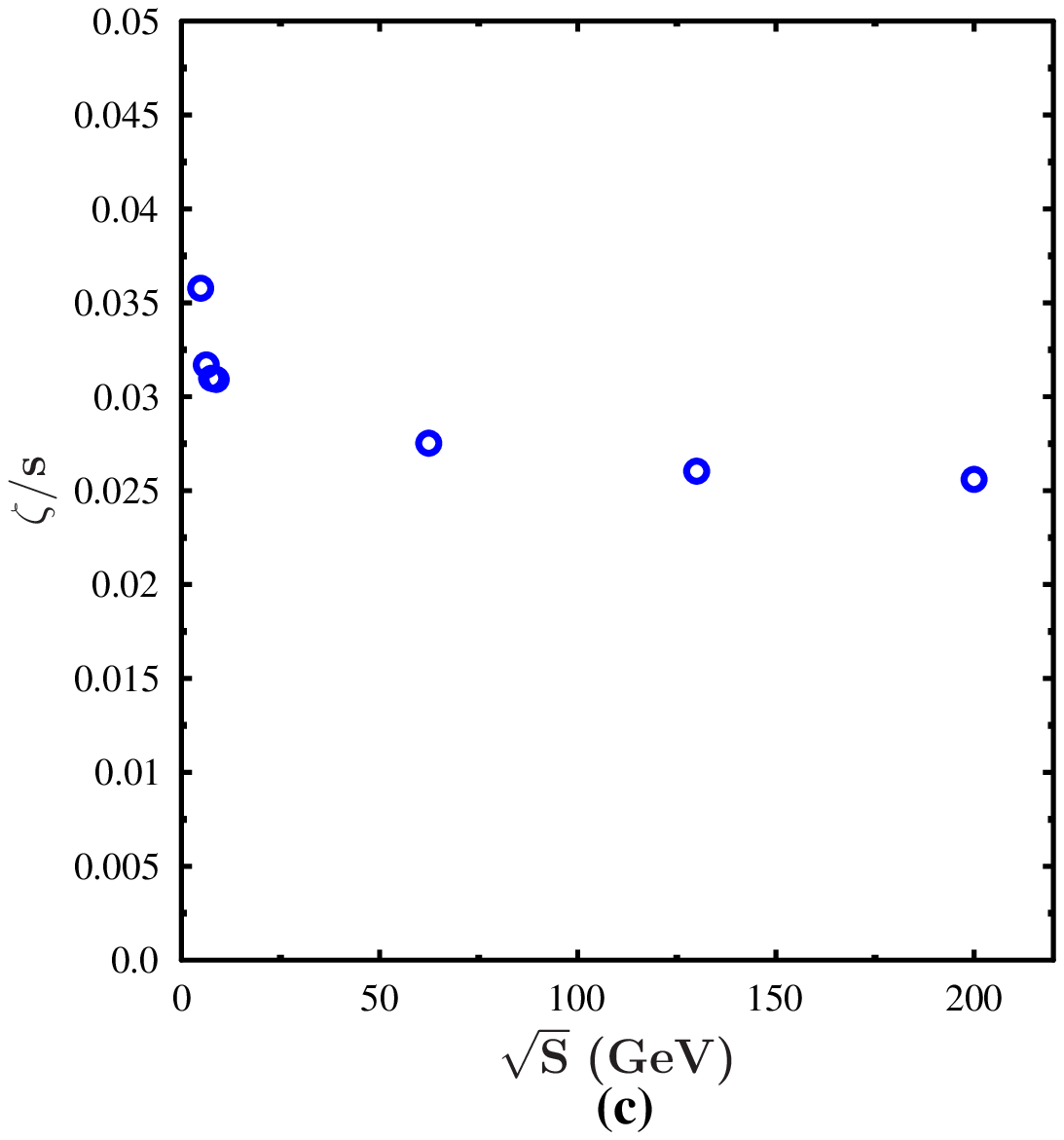}
  \caption{(Color online) Top panel shows $\eta/s$ along freezout line for $r_{h}=0.5$fm. Middle panel shows $\eta T/(\epsilon+P)$ along freezout line.  Bottom panel shows $\zeta/s$ along freezout line for $r_{h}=0.5$fm.} 
\label{snn}
 \end{figure}

 One can make connection with heavy ion collision experiments by finding the beam energy ($\sqrt{S}$) dependence of the temperature and chemical potential. This is extracted from a statistical thermal model description of the particle yield at various $\sqrt{S}$ \cite{cleymans}. The freeze out curve $T(\mu)$ is parametrized by$T(\mu)=a-b\mu^2-c\mu^4$, where, $a=0.166\pm0.002$ GeV, $b=0.139\pm0.016$ GeV$^{-1}$ and $c=0.053\pm0.021$ GeV$^{-3}$. The energy dependence of the baryon chemical potential is given as $\mu=d/(1+e\sqrt{S})$, with, $d=1.308\pm0.028$ GeV, and $e=0.273\pm0.008$ GeV$^{-1}$. From Fig.(\ref{snn}a) we observe that ratio $\eta/s$ is well above KSS bound at low center of mass energy and increases monotonically  to become constant at higher $\sqrt{S}$ along freezout curve. This is legitimate since low $\sqrt{S}$ corresponds to low temperature and high chemical potential along freezout curve at which shear viscosity is smaller. Fig.(\ref{snn}b) shows ratio $\eta T/(\epsilon+P)$ along chemical freezout. We observe that this ratio again remains constant apart from initial rise. Since ratio $\eta T/(\epsilon+P)$ is a true measure of fluidity at finite baryon chemical potential, we conclude that in chemical freezout transition the fluid behavior of hadron gas does not change\cite{jeon}.  Further, along freezout curve ratio $\zeta/s$ decreases monotonically first and then becomes independent at higher center of mass energies  as shown in Fig.(\ref{snn}c).

 \section{Summary and Conclusion}
 In this work we have estimated dissipative properties of hot and dense hadronic matter using Boltzmann equation in relaxation time approximation within ambit of excluded volume hadron resonance gas model. We assumed uniform hard core radius to all hadrons and included all hadrons and their resonances with mass cut-off 2.25 GeV. We considered only elastic scattering processes to compute transport coefficients. We find that shear viscosity to entropy ratio decreases with temperature.  Further at finite chemical potential $\eta/s$ shows same behavior as a function of temperature but ratio is smaller as compared to $\mu=0$. This decrease is solely due to rapid increase in entropy density at finite $\mu$. As pointed out in Ref.\cite{liao}, at finite baryon density it is $\eta T/(\epsilon+P)$ and not $\eta/s$ as a correct measure of fluidity. We find that effect of finite $\mu$ is more pronounced for $\eta T/(\epsilon+P)$ and this is again attributed to rapid rise in enthalpy. Further the bulk viscosity to entropy ratio decreases with temperature but at higher baryon chemical
potential this ratio is higher as compared to zero chemical potential. This is due to heavier baryonic excitations
which makes additional contribution to the bulk viscosity at finite baryon chemical potential. Further, along chemical freezout curve both the ratios $\eta/s$ and $\eta T/(\epsilon+P)$ remains constant apart from initial rise. This suggest that fluid behavior of hadron gas does not change along chemical freezout transition. Further, initially along along freezout line ratio $\zeta/s$ decreases monotonically and then becomes independent of central of mass energy . 
 
 In this work we have taken zero temperature hadron masses independent of temperature (T) and chemical potential ($\mu$) in the partition function. But hadron masses depends on T and $\mu$ since constituent quark masses (which makes large contribution to hadron mass) arises due to interaction with chiral condensates. These chiral condensates are depends on T and $\mu$ and actually melts as the temperature and chemical potential increases. So it would be more realistic to include T and $\mu$ dependent hadron masses in HRG model and to see the thermodynamics of such system and its effect on transport properties of hadronic matter. Work in this direction is in progress and will be reported elsewhere\cite{gurunxt}. 
 
\section*{ACKNOWLEDGMENTS}
G.P.K acknowledge Sean Gavin, Sangyong Jeon and Juan Torres-Rincon  for making comments on the manuscript. G.P.K also acknowledge G. Denicol, A.S. Khvorostukhin and S. Ghosh for providing numerical data.

 \def\karschkharzeev{F. Karsch, D. Kharzeev, and K. Tuchin, Phys. Lett. B
{\bf 663}, 217 (2008).}
\def\joglekar{J.C. Collins, A. Duncan, S.D. Joglekar, Phys. Rev. D {\bf 16}, 
438 (1977).}
\def\blaschke{J. Jankowski, D. Blaschke, M.Spalinski, Phys.Rev.D {\bf 87}, 105018
(2013). }
\def\gorenstein{M. Gorenstein, M. Hauer, O. Moroz, Phys.Rev.C {\bf 77}, 024911 (2008)}
\def\bugaev{K. Bugaev et al, Eur.Phys.J. A {\bf 49}, 30 (2013)}
\def\cpsingh{S.K. Tiwari, P.K. Srivastava, C.P. Singh, Phys.Rev. C {\bf 85},
014908 (2012) }
\def\chen{J.-W. Chen, Y-H. Li, Y.-F. Liu, and E. Nakano, Phys. Rev. D {\bf 76},
114011 (2007)}
\def\chennakano{J.-W. Chen, and E. Nakano, Phys. Lett. B {\bf 647}, 371 (2007)}
\def\itakura{K. Itakura, O. Morimatsu, and H. Otomo, Phys. Rev. D {\bf 77}, 014014
(2008)}
\def\cleymans{J. Cleymans, H. Oeschler, K. Redlich, and S. Wheaton, Phys.
Rev. C {\bf 73}, 034905 (2006)}
\def\hirano{P. Romatschke and U. Romatschke, Phys. Rev. Lett. {\bf 99}, 172301 (2007); T. Hirano and M. Gyulassy, Nucl. Phys. {\bf A 769}, 71, (2006).} 
\def\kss{P. Kovtun, D.T. Son and A.O. Starinets, Phys. Rev. Lett. {\bf 94},
 111601 (2005).}
\def\sasakiqp{C. Sasaki and K.Redlich,{\PRC{79}{055207}{2009}}.}
\def\sasakinjl{C. Sasaki and K.Redlich,{\NPA{832}{62}{2010}}.}
\def\ellislet{I.A. Shushpanov, J. Kapusta and P.J. Ellis,{\PRC{59}{2931}{1999}}
; P.J. Ellis, J.I. Kapusta, H.-B. Tang,{\PLB{443}{63}{1998}}.}
\def\prakashwiranata{A. Wiranata and Madappa Prakash, Phys. Rev. C {\bf 85}, 054908 (2012).}
\def\purnendu{P. Chakraborty and J.I. Kapusta {\PRC{83}{014906}{2011}}.}
\def\greco{S.Plumari,A. Paglisi,F. Scardina and V. Greco,{\PRC{83}{054902}{2012}.}}
\def\bes{H. Caines, arXiv:0906.0305 [nucl-ex], 2009.}
\def\greinerprl{J. Noronha-Hostler,J. Noronha and C. Greiner, 
{\PRL{103}{172302}{2009}}.}
\def\greinerprc{J. Noronha-Hostler,J. Noronha and C. Greiner
, {\PRC{86}{024913}{2012}}.}
\def\igorgreiner{J. Noronha-Hostler, C. Greiner and I. Shovkovy,
, {\PRL{100}{252301}{2008}}.}
\def\nakano{J.W. Chen,Y.H. Li, Y.F. Liu and E. Nakano,
 {\PRD{76}{114011}{2007}}.}
\def\itakura{K. Itakura, O. Morimatsu, H. Otomo, {\PRD{77}{014014}{2008}}.}
\def\wang{M.Wang,Y. Jiang, B. Wang, W. Sun and H. Zong, Mod. Phys. lett.
{\bf A76}, 1797,(2011).}
\def\hrgexp{P. Braunmunzinger, J. Stachel, J.P. Wessels and N. Xu,
{\PLB{365}{1}{1996}}; G.D. Yen and M.I. Gorenstein, {\PRC{59}{2788}{1999}};
F. Becattini, J. Cleymans, A. Keranen, E. suhonen and K. Redlich, 
{\PRC{64}{024901}{2001}}.}
\def\rischkegorenstein{.D.H. Rischke, M.I. Gorenstein, H. Stoecker and
W. Greiner, Z.Phys. C {\bf 51}, 485 (1991).}
\def\hmnjl{Amruta Mishra and Hiranmaya Mishra, {\PRD{74}{054024}{2006}}.}
\def\pdgb{C. Amseler {\it et al}, {\PLB{667}{1}{2008}}.}
\def\shuryak{E.V. Shuryak, Yad. Fiz. {\bf 16},395, (1972).}
\def\leupold{S. Leupold, J. Phys. G{\bf32},2199,(2006)}
\def\peter{A. Andronic, P. Braun-Munzinger , J. Stachel and M. Winn,
{\PLB{718}{80}{2012}}}
\def\blum{M. Blum, B. Kamfer, R. Schluze, D. Seipt and U. Heinz,{\PRC{76}{034901}{2007}}.}
\def\jaminplb{M. Jamin{\PLB{538}{71}{2002}}.}
\def\ghosh{Sabyasachi Ghosh{\PRC{90}{025202}{2014}}.}
\def\csernai{L.P. Csernai, J.I. Kapusta and L.D. McLerran,{\PRL{97}{152303}{2006}}.}
\def\hagedorn{R. Hagedorn, Nuovo Cim. Suppl. 3,147 (1965); Nuovo Sim. A56,1027 (1968).}
\def\torieri{G. Torrieri and I. Mishustin,{\PRC{77}{034903}{2008}}.}
\def\fernandez{D. Fernandiz-Fraile and A.G. Nicola,{\PRL{102}{121601}{2009}}.}
\def\caron{S.Caron,{\PRD{79}{125009}{2009}}.}
\def\latticemeyer{H.B. Meyer,{\PRL{100}{162001}{2008}}.} 
\def\romatschke{P.Romatscke and D.T. Son,{\PRD{80}{065021}{2009}}.}
\def\moore{G.D. Mooore and O. Sarem, J. High Energy Phys. JHEP0809(2008)015.}
\def\dobado{A.Dobado and J. M. Torres-Rincon {\PRD{86}{074021}{2012}}.}
\def\daniel{D. Fernandez-Fraile, {\PRD{83}{065001}{2011}}.}
\def\gurunpa{Guru Prakash Kadam, H. Mishra, Nuclear Physics A {\bf 934}, (2015) 133–147.}
\def\gurumpla{Guru Prakash Kadam, Mod.Phys.Lett. A {\bf 30}, (2015) 10, 1550031.}
\def\demir{N. Demir, S.A. Bass, Phys. Rev. Lett. {\bf 102}, (2009) 172302.}
\def\phsd{V. Ozvenchuk, O. Linnyk, M.I. Gorenstein, E.L. Bratkovskaya, W. Cassing, Phys. Rev. C {\bf 87}, (2013) 064903.}
\def\ehrgrishke{D. H. Rischke, M. I. Gorenstein, H. Sto ̈cker, and W. Greiner,
Z. Phys. C {\bf 51}, 485 (1991).͒}
\def\ehrgclaymans{J. Cleymans, M. I. Gorenstein, J. Stalnacke, and E. Suhonen,
Phys. Scr. {\bf 48}, 277 (1993͒).}
\def\gavin{S. Gavin,  Nucl.Phys. A {\bf 435}, 826 (1985).}
\def\cannoni{Mirco Cannoni, Phys. Rev. D {\bf 89}, 103533 (2014).}
\def\gelmini{P. Gondolo and G. Gelmini, Nucl. Phys. B {\bf 360}, 145 (1991).}
\def\jeon{G.S. Denicol, C. Gale,  S. Jeon,  and J. Noronha, Phys. Rev. C {\bf 88}, 064901 (2013).}
\def\sinha{A. Buchel, R. C. Myers and A. Sinha, JHEP {\bf 0903}, 084 (2009).}
\def\heppe{P. Braun-Munzinger, I. Heppe, J. Stachel, Phys. Lett. B {\bf 465}, 15 (1999).}
\def\bugaev{K.A. Bugaev, D.R. Oliinychenko, A.S. Sorin, and G.M. Zinovjev, arxive:1208.5968v1.}
\def\cleymans{J. Cleymans, H. Oeschler, K. Redlich, S. Wheaton, Phys. Rev. C {\bf 73}, 034905 (2006).}
\def\moore{E. Lu, G. D. Moore, Phys. Rev. C {\bf 83}, 044901 (2011).}
\def\amseler{C. Amseler, et al., Phys. Lett. B {\bf 667}, 1 (2008).}
\def\moroz{O. Moroz, Ukr.J.Phys. {\bf 58}, 1127 (2013).}
\def\gurunxt{Guru Prakash Kadam, H. Mishra, In preparation.}
\def\yen{G. D. Yen, M. I. Gorenstein, W. Greiner, and S. N. Yang, Phys.Rev. C {\bf 56}, 2210 (1997).}
\def\sarkarghosh{ S. Ghosh, G. Krein, S. Sarkar, Phys.Rev. C {\bf 89}, 045201 (2014).}
\def\finazzo{ S. I. Finazzo, R. Rougemont, H. Marrochio, J. Noronha, JHEP {\bf 1502}, 051 (2015).}
\def\khvorostukhin{A.S. Khvorostukhin, V.D. Toneev, D.N. Voskresensky, Nucl. Phys. A {\bf 845}, 106 (2010).}
\def\sghosh{S. Ghosh, Int. J. Mod. Phys. A {\bf 29}, 1450054 (2014).}
\def\sghoshnucl{S. Ghosh, Phys. Rev. C {\bf 90}, 025202 (2014).}
\def\sghoshnjl{S. Ghosh,  A. Lahiri, S. Majumder, R. Ray, S. K. Ghosh, Phys. Rev. C {\bf 88}, 068201 (2013).}
\def\cohen{ T.D. Cohen, Phys. Rev. Lett. {\bf 99}, 021602 (2007).}
\def\rebhan{ A. Rebhan and D. Steineder, Phys. Rev. Lett. {\bf 108}, 021601 (2012).}
\def\mamo{ K.A. Mamo, JHEP {\bf 70}, 1210 (2012).}
\def\weise{R. Lang, N. Kaiser, and W. Weise, arxive:1506.02459}
\def\dobadojuan{A. Dobado, F. J. Llanes-Estrada and J. M. Torres-Rincon, Phys. Lett. B {\bf 702}, 43 (2011).}
\def\gyulassy{P. Danielewicz and M. Gyulassy, Phys.Rev. D {\bf 31}, 53 (1985).}
\def\borsonyi{S. Borsonyi, et al., JHEP {\bf 1011}, 077 (2010).}
\def\hotQCD{Bazavov et al., Phys. Rev. D {\bf 90}, 094503 (2014).}
\def\WPblk{A. Wiranata and M. Prakash, Nucl. Phys. A {\bf 830}, 219–222 (2009).}
\def\prakash{M. Prakash, M. Prakash, R. Venugopalan and G. Welke, Phys.Rept. {\bf 227}, 321-366 (1993).}
\def\wiranatakoch{A. Wiranata, V. Koch and  M. Prakash, X.N. Wang, J.Phys.Conf.Ser. {\bf 509}, 012049 (2014).}
\def\wiranataprc{A. Wiranata and M. Prakash, Phys.Rev. C {\bf 85}, 054908 (2012).}
\def\wiraprapurn{ A. Wiranata, M. Prakash and  P. Chakraborty, Central Eur.J.Phys. {\bf 10}, 1349-1351 (2012).}
\def\wiranatademir{N. Demir and A. Wiranata, J.Phys.Conf.Ser. {\bf 535}, 012018 (2014).}
\def\gorensteinplb{M. Gorenstein, V. Petrov and G. Zinovjev, Phys.Lett. B {\bf 106}, 327-330 (1981).}
\def\goregranddonprc{G. D. Yen, M. Gorenstein, W. Greiner and Shin Nan Yang, Phys. Rev. C {\bf 56}, 2210 (1997).}
\def\gorensteinjpg{M. Gorenstein, H. St\"{o}ker, G. D. Yen, Shin Nan Yang and W. Greiner, J. Phys. G {\bf 24}, 1777 (1998).}
\def\yengorenprc{G.D. Yen and M. Gorenstein, Phys. Rev. C {\bf 59}, 2788 (2009).}
\def\gorengreinerjpg{M. Gorenstein, W. Greiner and Shin Nan Yang, J. Phys. G {\bf 24}, 725 (1998).}
\def\gorengreinerprc{M. Gorenstein, M. Ga'zdicki and W. Greiner, Phys. Rev. C {\bf 72}, 024909 (2005).}
\def\tawfik{A. Tawfik and M. Wahba, Ann. Phys. {\bf 522}, 849-856 (2010).}
\def\jeonyaffe{S. Jeon and L. Yaffe, Phys.Rev. D {\bf 53}, 5799-5809 (1996).}
\def\liao{J. Liao, V. Koch, Phys. Rev. D {\bf 81}, 014902 (2010).}
\def\Khov{A. Khvorostukhin, V. Toneev, D. Voskresensky, Nucl. Phys. A {\bf 845}, 106–146, (2010).}
\def\nicola{D. Fernandez-Fraile, A. Gomez-Nicola, Eur.Phys.J. C {\bf 62}, 37-54 (2009).}

\end{document}